\documentclass[structabstract,english]{aa}

\usepackage{txfonts}
\usepackage{amsfonts}
\usepackage{amstext}
\usepackage{textcomp}
\usepackage{natbib}
\usepackage{wasysym}
\usepackage{graphicx}
\usepackage{babel}
\usepackage{color}

\makeatletter

\usepackage{hyperref}
\hypersetup{
    colorlinks,%
    citecolor=blue,%
    filecolor=blue,%
    linkcolor=blue,%
    urlcolor=blue
}
\usepackage[all]{hypcap}

\makeatother

\newcommand{\rmd}{{\rm d}}
\newcommand{\bk}{{\mathbf{k}}}
\newcommand{\bx}{{\mathbf{x}}}
\newcommand{\BR}{{\mathbb{R}}}
\newcommand{\CB}{{\cal B}}
\newcommand{\cD}{{\cal D}}
\newcommand{\CQ}{{\cal Q}}
\newcommand{\CR}{{\cal R}}
\newcommand{\CO}{{\cal O}}
\newcommand{\average}[1]{\left\langle #1 \right\rangle_\cD}
\newcommand{\av}[1]{\left\langle #1 \right\rangle}

\begin{document}

\title{Inhomogeneity-induced variance of cosmological parameters}
\titlerunning{Inhomogeneity-induced variance of cosmological parameters}

\author{Alexander Wiegand and Dominik J. Schwarz}
\authorrunning{A. Wiegand, D.J. Schwarz}

\institute{Fakult\"at f\"ur Physik, Universit\"at Bielefeld, 
Universit\"atsstrasse 25, D--33615 Bielefeld, Germany\\
\email{wiegand and dschwarz both at physik dot uni-bielefeld dot de}}

\begin{abstract}
{Modern cosmology relies on the assumption of large-scale isotropy and homogeneity of 
the Universe. However, locally the Universe is inhomogeneous and anisotropic. This raises the question of how 
local measurements (at the $\sim 10^2$ Mpc scale) can be used to determine the global cosmological 
parameters (defined at the $\sim 10^4$ Mpc scale)?}
{We connect the questions of cosmological backreaction, cosmic averaging and the estimation 
of cosmological parameters and show how they relate to the problem of cosmic variance.}
{We used Buchert's averaging formalism and determined a set of locally averaged cosmological 
parameters in the context of the flat $\Lambda$ cold dark matter model. We calculated their 
ensemble means (i.e.~their global value) and variances (i.e.~their cosmic variance). 
We applied our results to typical survey geometries and focused on the study of the effects of 
local fluctuations of the curvature parameter.}
{We show that in the context of standard cosmology at large scales (larger than the homogeneity 
scale and in the linear regime), the question of cosmological backreaction and averaging can be 
reformulated as the question of cosmic variance. The cosmic variance is found to be highest in the 
curvature parameter. We propose to use the observed variance of cosmological parameters 
to measure the growth factor.}
{Cosmological backreaction and averaging are real effects that have been measured 
already for a long time, e.g.~by the fluctuations of the matter density contrast averaged over 
spheres of a certain radius.  Backreaction and averaging effects from scales in the linear 
regime, as considered in this work, are shown to be important for the precise measurement of 
cosmological parameters.}
\end{abstract}

\keywords{Cosmology:theory, cosmological parameters, large-scale structure of Universe}

\maketitle

\section{Introduction}

How do inhomogeneities in the matter distribution of the Universe affect 
our conception of its expansion history and our ability to measure 
cosmological parameters? Typically, these measurements rely on the
averaging of a large number of individual observations. In an idealised 
situation we can think of them as volume averages. To give an example,
the power spectrum, which is the Fourier--transformed two--point correlation function, 
may be seen as a volume average with weight $e^{i k x}$. Measurements of the 
properties of the large--scale structure rely on the observation of large volumes that have 
been pushed forward to ever higher redshifts in the last decade, from the two--degree 
Field survey (2dF, \citealt{survey:2dF}) over the Sloan Digital Sky Survey 
(SDSS, \citealt{survey:SDSS}) to the current WiggleZ \citep{survey:WiggleZ} and 
Baryon Oscillation Spectroscopic Survey (BOSS, \citealt{survey:SDSSIII}).

A theorem by Buchert states that the evolution of any volume-averaged 
comoving domain of an arbitrary irrotational dust Universe may be 
described  by the equations of a Friedmann--Lema\^itre--Robertson--Walker (FLRW) model, but driven by effective sources that encode inhomogeneities \citep{buchert:dust,buchert:fluid}. 
The consequences of this have been studied extensively in 
perturbation theory \citep{kolb:perturb,li:onset,li:scale,brown1,brown2,larena,clarkson:pert,2011ToBePub}
and for non-perturbative models \citep{morphon,kolb:swisscheese,rasanen:peakmodel,meatball,roy:chaplygin}. 
Apart from the ongoing debate to what extent the global evolution is modified
through backreaction effects from small-scale inhomogeneities 
\citep{rasanen:pertsecorder,ishi-wald,kolb:de,review,curvestim,wiegand}, 
\cite{li:scale} showed that the measurement of cosmological parameters is limited 
by uncertainties concerning the relation between observable locally and
unobservable globally averaged quantities. 

In contrast to the well--studied cosmic variance of the cosmic 
microwave background, which is most relevant at the largest angular scales,
the theoretical limitation on our ability to predict observations at low 
redshift arises not only from the fact that we observe only one  
Universe, but also from the fact that we sample a finite domain 
(much smaller than the Hubble volume). Both limitations contribute to
the cosmic variance. This is different from the sampling variance caused by shot
noise, i.e.~the limitation of the sampling of a particular domain because of the finite number of supernovae (SN) or galaxies observed. In the era of precision cosmology, the errors caused by 
cosmic variance may become a major component of the error budget.

The purpose of this work is to demonstrate that the questions of cosmic
averaging, cosmological backreaction, and the problems of cosmic variance
and parameter estimation are closely linked. We demonstrate that
cosmic variance is actually one of the aspects of cosmological
backreaction.

There have already been many studies
on the effect of the local clumpiness on our ability to measure cosmological
parameters. However, they focused mainly on the fluctuations in
the matter density and on the variance of the Hubble
rate. The former question has gained renewed interest in view of 
deep redshift surveys such as GOODS \citep{survey:goods}, 
GEMS \citep{survey:gems} or COSMOS \citep{survey:cosmos}.
Because the considered survey fields are small, the variance of the 
matter density is an important ingredient in the error budget and it has 
been found to be in the range of 20\% and more, 
demonstrated empirically in SDSS data by 
\cite{driver:cosvar} and calculated numerically from linear perturbation
theory in \cite{moster:cookbook}.

The variance of the Hubble rate has been considered in the setup 
of this work, i.e.~first order linear perturbation theory in 
comoving synchronous gauge, in \cite{li:scale}. 
A calculation of the same effect in Newtonian gauge
has been performed in \cite{clarkson:pert} and \cite{clarkson:hubble}; 
the two methods agree.
Calculations of the fluctuations in the Hubble rate owing to peculiar
velocities alone have a longer history 
\citep{hubble:kaiser,hubble:turner,hubble:shiturner,hubble:wangturner}.
The effects of inhomogeneities on other cosmic parameters have been studied in Newtonian cosmology \citep{bks:basics,bks:main}.

There has been less activity in studying the effects of the third important 
player besides number density and Hubble expansion: cosmic
curvature. Even if the Universe is spatially flat, as expected by the
scenario of cosmological inflation and consistent with the observations
of the temperature anisotropies of the cosmic microwave background, 
the local curvature may be quite different. To answer the question how 
big this difference actually is for realistic survey volumes, we here extend 
the analysis of \cite{li:onset,li:scale}, where these effects have
been estimated for the first time. For the spatially flat Einstein-de Sitter 
(EdS) model, the averaged curvature parameter $\Omega_{\cal R}^{\cal D}$ 
has been shown to deviate from zero by $\sim 0.1$ on domains at the 
$100$~Mpc scale.

Here we adapt the analysis of \cite{li:scale} to the case of a $\Lambda$CDM
universe, introduce a more realistic power spectrum and use observationally
interesting window functions, not restricted to full--sky measurements.

\begin{figure}
\includegraphics[width=\linewidth]{%
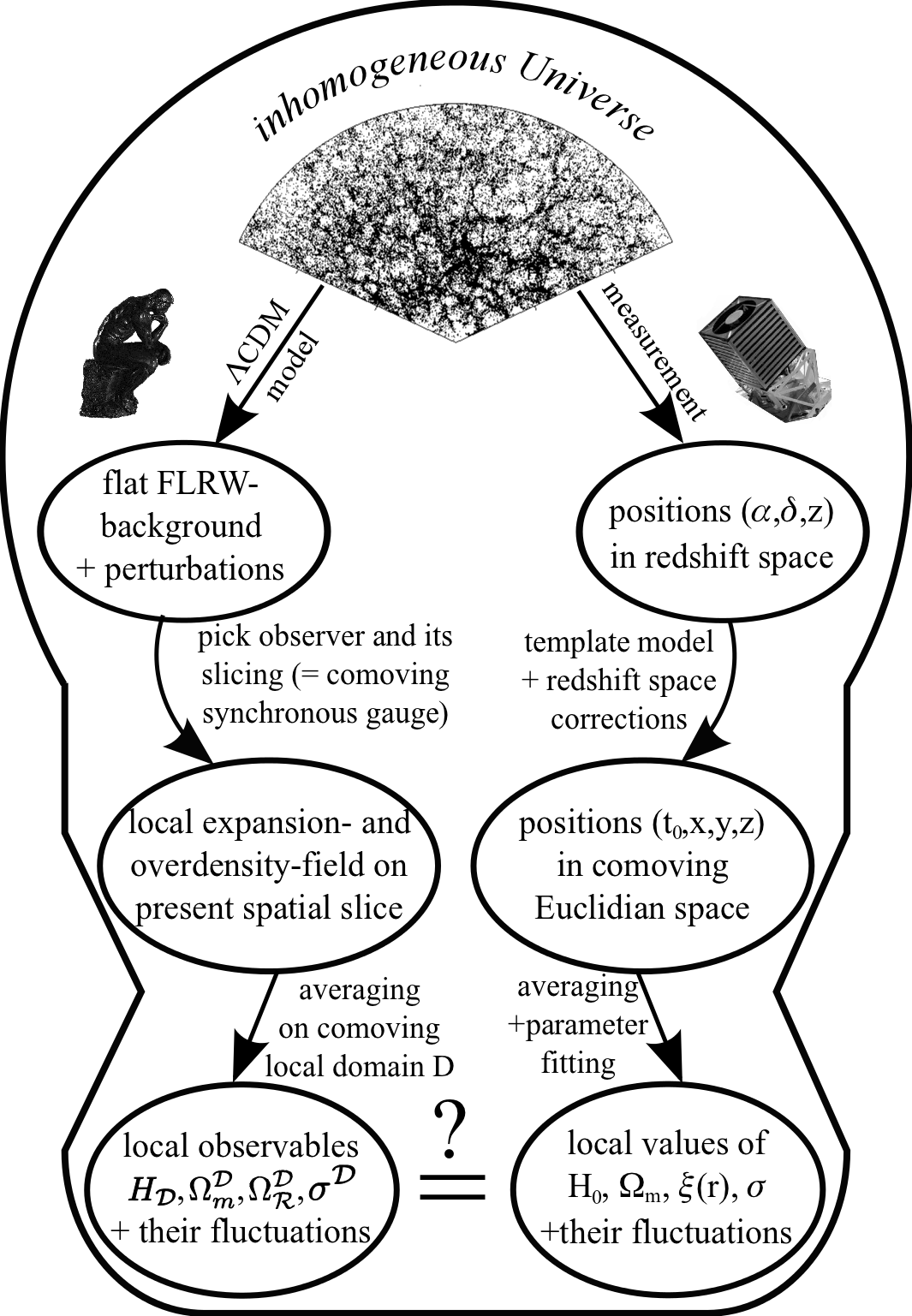}
\caption{Comparison of the theorist's and observer's view on the Universe.
Our calculation in comoving synchronous gauge facilitates the description of
the boundaries of the experimentally investigated regions in our Universe.
Note, that recently there have been attempts by \cite{bonvinDurrer} to directly relate the two upper circles. This was done by calculating the predictions for the quantities in redshift space explicitly from the perturbed $\Lambda$CDM model.
\label{fig:theory-measurement}}
\end{figure}

There has been quite some confusion about the choice of
gauge and the dependence of the averaged quantities on it. Recently
it has been shown \citep{veneziano} that this is not a problem if one
consistently works in one gauge and then expresses the
quantity that is finally observed in this frame as well. This is easier
in some gauges than in others, but the result is (as expected) 
the same, as is also
confirmed explicitly by the fact that our results are consistent
with those of \cite{clarkson:pert} and \cite{clarkson:hubble}, obtained 
in Newtonian gauge.
The quest for simplicity explains our choice to use comoving synchronous
gauge, because this is the frame that is closest to the one used by the
observers.

Fig.~\ref{fig:theory-measurement} depicts the theorist's
and the observer's view on the Universe in a schematical way. It
points out that in the end it is the average quantities that we are
interested in, but that in an intermediate step, observers like to
think of the objects they measure to lie in a comoving space with
simple Euclidian distances. The comoving synchronous gauge, which has
a clear notion of "today" for a fluid observer sitting in a
galaxy as we do, helps to define the things observers measure in a simple way.

In Section~\ref{sec:Inhomogeneity-and-expansion}
we establish the conceptual framework for the study of effects
of inhomogeneities on observable quantities. Section~\ref{sec:Var-cosmic-pars} 
and \ref{sec:Var-loc-cosm-par}
generalize some of the results of \cite{li:onset,li:scale} and \cite{li:thesis} from
an Einstein--de Sitter (EdS) to a $\Lambda$CDM background and implement a more realistic
matter power spectrum. Section~\ref{sec:survey-geom} investigates
the effects of various window functions, again extending the analysis of 
\cite{li:scale,li:thesis}.
Section~\ref{sec:curv-fluct} concentrates on deriving the magnitude
of curvature fluctuations for realistic window functions. 
Section~\ref{sec:Variances-on-BAO} 
applies the formalism to the local distance
measure $D_{V}$, determined in the observation of baryonic acoustic
oscillations (BAO). Section~\ref{sec:Expansion-history}
is a remark on the link between the variance of averaged expansion rates
at different epochs and the background
evolution, before we conclude in Section~\ref{sec:Conclusion}.

\section{Inhomogeneity and expansion
\label{sec:Inhomogeneity-and-expansion}}

We assume that the overall evolution of the Universe is described by 
a flat $\Lambda$CDM model, which we adopt as our background model
throughout this work. Global spatial flatness does not prevent the 
local curvature to deviate from zero.

The distribution of nearby galaxies indicates that  
the scales at which the Universe is inhomogeneous reach out to at 
least 100 Mpc. Above these scales it is not yet established if 
there is a turnover to homogeneity, as was claimed in \cite{hogg:homo}, 
or if the correlations in the matter distribution merely become weaker, but persist up 
to larger scales, as is discussed in \cite{labini:inhom}. At least morphologically,
homogeneity has not been found up to scales of about 200 Mpc 
\citep{morph:pscz,morph:iras,morph:SDSS}.

Consequently, we need a formalism
that is applicable in the presence of inhomogeneities at least for the description of the local expansion. This may be accomplished considering 
spatial domains $\cD$ and averaging over their locally inhomogeneous
observables \citep{buchert:dust,buchert:fluid}. Technically one
performs a 3+1 split of spacetime. Because we
considered pressureless matter only, we chose a comoving foliation
in which the spatial hypersurfaces are orthogonal to the cosmic time.
This means that the formalism will not be able to take into account
lightcone effects. Therefore, it is well adapted for regions
of the universe that are not too extended, a notion described more precisely in Sect.~\ref{sec:survey-geom}.
The equations then describe the evolution of the
volume of the domain $\cD$, given by 
$\left|\cD\right|_{g}:=\int_{\cD}\, {\rm d}\mu_{g}$,
where 
${\rm d}\mu_{g}:= [{}^{(3)}g\left(t,\bx\right)]^{1/2}
{\rm d}^{3}x$ and ${}^{(3)}g$ is the fully inhomogeneous three--metric
of a spatial slice. 

To obtain an analogy to the standard Friedmann
equations, one defines an average scale factor from this volume
\begin{equation}
a_{\cD}\left(t\right):=\left(\frac{\left|\cD\right|_{g}}{\left|\cD_{0}\right|_{g}}\right)^{\frac{1}{3}}\;,
\label{eq:scale-factor}
\end{equation}
where the subindex $0$ denotes "today", as throughout the rest of this work. 
The definition implies $a_{\cD_{0}}=1$.
In analogy to the background model 
$H_{\cD}:=\dot{a}_{\cD}/a_{\cD}$.
The foliation may be used to define an average over a three--scalar observable $O$ on a domain $\cD$ in the spatial hypersurface
\begin{equation}
\average{O}\left(t\right) :=
\frac{\int_{\cD}\, O\left(t,\bx\right){\rm d}\mu_{g}}{\int_{\cD}\, {\rm d}\mu_{g}}.
\label{eq:Def-Mittel}
\end{equation}
Examples for these observables include the matter density $\varrho$ or the redshift $z$ of a group of galaxies in the domain $\cD$.

The scalar parts of Einstein's equations for an inhomogeneous matter
source become evolution equations for the volume scale factor driven by quantities determined by this average:
\begin{eqnarray}
3\frac{\ddot{a}_{\cD}}{a_{\cD}} & = & 
-4\pi G\average{\varrho}+\CQ_{\cD} + \Lambda,%
\label{eq:Raychaudhuri-Mittel}\\
3H_{\cD}^{2} & = & 
8\pi G\average{\varrho}-\frac{1}{2}\average{\CR}-\frac{1}{2}\CQ_{\cD}+\Lambda,%
\label{eq:Hamilton-Mittel}\\
0 & = & \partial_{t}\average{\varrho}+3H_{\cD}\average{\varrho}.%
\label{eq:Konservation-Mittel}
\end{eqnarray}
The expansion of the domain $\cD$ is determined
by the average matter density,
the cosmological constant, the average intrinsic scalar 
curvature $\average{\CR}$ and the kinematical 
backreaction $\CQ_{\cD}$. The latter encodes the departure
of the domain from a homogeneous distribution and is a linear combination
of the variance of the expansion rate and the variance of the shear
scalar.

Equations (\ref{eq:Raychaudhuri-Mittel}) to (\ref{eq:Konservation-Mittel}) 
mean that the local evolution of any inhomogeneous domain
is described by a set of equations that corresponds to the Friedmann
equations.

The cosmic parameters, defined by 
\begin{eqnarray}
\Omega_{m}^{\cD} & := & \frac{8\pi G}{3H_{\cD}^{2}}\average{\varrho}, \quad
\Omega_{\Lambda}^{\cD}:=\frac{\Lambda}{3H_{\cD}^{2}},
\label{eq:Def-Omega-Parameter}\\
\Omega_{\CR}^{\cD} & := & -\frac{\average{\CR}}{6H_{\cD}^{2}}, \quad
\Omega_{\CQ}^{\cD}:=-\frac{\CQ_{\cD}}{6H_{\cD}^{2}}, \nonumber
\end{eqnarray}
are domain--dependent. Owing to the fluctuating matter density,
the curvature and the average local expansion rate will also fluctuate.
When we constrain ourselves to the
perturbative regime, the modification due to $\CQ_{\cD}$ is important on 
scales of the order of $10$ Mpc \citep{li:scale}. 

Here we are interested to see to what extent the values of the
parameters (\ref{eq:Def-Omega-Parameter}) vary if we look at different
domains of size $\cD$ in the Universe. 
We could therefore define an average over many equivalent domains $\cD$ at different locations in our current spatial slice.
For an ergodic
process, however, this is the same as the variance of an ensemble average over
many realizations of the Universe keeping the domain $\cD$ fixed, but changing the initial conditions of the matter distribution. This is the quantity that we calculate in theory and therefore we have to rely on the assumption of ergodicity when comparing our results with the observation.
In our case this ensemble
average is taken over quantities that are volume averages. This means
that for any observable $O$ there are two different averages involved. 
The domain averaging, $\average{O}$, and the ensemble average, $\overline{O}$.
We assume that both averaging procedures commute. 

The fluctuations are then characterized by the variance
with respect to the ensemble averaging process,
\begin{equation}
\sigma\left(\average{O}\right):=\left(\overline{\average{O}^{2}}-\overline{\average{O}}^{2}\right)^{\frac{1}{2}}\;.
\label{eq:var}
\end{equation}
An example for a common observable calculated in this manner would be $\sigma_{8}$ as the ensemble r.m.s.~fluctuation of the matter density field.
To quantify these fluctuations of a general observable $O$, we use the theory of cosmological
perturbations. In \cite{li:onset} it has been shown that to linear order,
the $\CQ_{\cD}$--term in equations (\ref{eq:Hamilton-Mittel})
and (\ref{eq:Raychaudhuri-Mittel}) vanishes. $\average{\varrho}$,
$\average{\CR}$ and $H_{\cD}$ however, have linear corrections.

Furthermore, \cite{li:thesis} argued that there is no second--order 
contribution to the fluctuations for Gaussian density perturbations
if their linear contributions are finite.
This may be seen by decomposing the observable $O$ into successive orders 
$O = O^{\left(0\right)} + O^{\left(1\right)} + O^{\left(2\right)}+\cdots$.
It is usually assumed that $\overline{O^{(1)}} = 0$. Now, for
Gaussian perturbations only terms of even order give
rise to non-trivial contributions. Therefore, (\ref{eq:var}) may be 
expressed as 
\begin{eqnarray}
\sigma\left(O\right)=\sqrt{\overline{\left(O^{\left(1\right)}\right)^{2}}}
\left(1+\frac{\overline{\left(O^{\left(2\right)}\right)^{2}}-
\left(\overline{O^{\left(2\right)}}\right)^{2}+
2\overline{O^{\left(1\right)}O^{\left(3\right)}}}{2\overline{\left(O^{\left(1\right)}\right)^{2}}}\right),
\label{var1}
\end{eqnarray}
which shows that the correction to the leading order linear term is
already of third order. This is why we content ourselves for the evaluation
of the fluctuations in the parameters $\average{\varrho}$, $\average{\CR}$
and $H_{\cD}$ or $\Omega_{m}^{\cD}$, $\Omega_{\CR}^{\cD}$ and $H_{\cD}$
to a first order treatment. This argument does not apply if 
$\overline{\left(O^{\left(1\right)}\right)^{2}} = 0$, as is the case for $\CQ_\cD$. In that 
case $\overline{\CQ_\cD}$ and $\sigma(\CQ_\cD)$ are of second order in perturbation
theory.

\section{Cosmological parameters and their mean from local averaging
\label{sec:Var-cosmic-pars}}

The analysis of this work is based on standard perturbation theory
in comoving (synchronous) gauge and we use results and notation
of \citet{li:onset}. The perturbed line element 
\begin{equation}
\mbox{d}s^{2}=a^{2}\left(\eta\right)\left\{ -\mbox{d}\eta^{2}+\left[\left(1-2\psi^{\left(1\right)}\right)\delta_{ij}+D_{ij}\chi^{\left(1\right)}\right]\rmd x^{i}\rmd x^{j}\right\} 
\end{equation}
 defines the metric potentials $\psi^{(1)}(\eta,{\bf x})$ and $\chi^{(1)}(\eta,{\bf x})$. Below
we use the convention $a_{0}=1$ for today's scale factor. We use conformal
time $\eta$ and the traceless differential operator $D_{ij}=\partial_{i}\partial_{j}-\frac{1}{3}\delta_{ij}\Delta$
on a spatially flat background. The geometrical quantities of interest
are the local expansion rate and the spatial curvature. The former
follows from the expansion tensor and reads
\begin{equation}
\theta=\frac{3}{a}\left(\frac{a'}{a}-{\psi^{\left(1\right)}}'\right),\label{eq:theta-pert}
\end{equation}
where $()^{\prime}$ stands for the derivative with respect to conformal
time. Calculating the spatial Ricci curvature from the above metric yields
\begin{equation}
\CR=\frac{12}{a^{2}}\left(2\frac{a'}{a}{\psi^{\left(1\right)}}'+{\psi^{\left(1\right)}}''\right).
\label{eq:R-pert}
\end{equation}
By the covariant conservation of the energy momentum tensor, 
$\psi^{\left(1\right)}$
is related to the matter density contrast
\begin{equation}
\delta\left(\eta,\bx\right) := {\rho^{\left(1\right)}\over\rho^{\left(0\right)}}
\end{equation} 
by 
\begin{equation}
\psi^{\left(1\right)}=\frac{1}{3}\delta-\bar{\zeta}\left({\bf x}\right),\label{eq:def-psi}
\end{equation}
with $\bar{\zeta}\left({\bf x}\right)$ denoting a constant of integration.
This constant plays no role in the following, because $\theta$ and ${\cal R}$ 
involve only time derivatives of $\psi^{\left(1\right)}$.

For dust and a cosmological constant, Einstein's equations give the well--known relation
\begin{equation}
\delta''+\frac{a'}{a}\delta'=\frac{4\pi G\rho_{0}^{\left(0\right)}}{a}\delta,
\end{equation}
at first order in perturbation theory. For the $\Lambda$CDM model 
the solution reads [see, e.g. \citet{green2005}]
\begin{equation}
\delta\left(a,{\bf x}\right)
= \frac{D\left(a\right)}{D\left(1\right)}\delta_{0} ({\bf x}),
\end{equation}
where $\delta_{0}\left(\bx\right)$ is the density perturbation today.
$D\left(a\right)$ is the growth factor given by
\begin{equation}
D\left(a\right)=a\,_{2}F_{1}\left(1,\frac{1}{3};\frac{11}{6};-ca^{3}\right),\quad \mbox{with\ }
c\equiv {\Omega_{\Lambda}\over \Omega_{m}}
\end{equation}
and $_{2}F_{1}$ is a
hypergeometric function. In the following we denote today's value of 
the growth factor by $D_0  \equiv D\left(1\right)$.

Plugging this solution into (\ref{eq:theta-pert})
and using (\ref{eq:def-psi}), we find the local expansion rate
\begin{equation}
\frac{1}{3}\theta(a,\bx)
=
H_{0}\sqrt{\frac{\Omega_{m}}{a^{3}}}\sqrt{1+ca^{3}}
\left(1-\frac{1}{3}f\left(a\right)\delta\left(a,\bx\right)\right),%
\label{eq:thetapert}
\end{equation}
expressed in terms of the growth rate 
\begin{equation}
f\left(a\right) := \frac{\rmd \ln D\left(a\right)}{\rmd \ln a} =
\frac{5\frac{a}{D\left(a\right)}-3}{2\left(1+ca^{3}\right)}.
\label{eq:growthrate}
\end{equation}
From (\ref{eq:R-pert}) we find the local spatial curvature
\begin{equation}
\CR (a,\bx) = 
10\frac{1}{a^{2}}H_{\text{0}}^{2}\Omega_{m}\frac{\delta_{0}(\bx)}{D_{0}}.
\end{equation}

From these quantities we can define local $\Omega$ functions,
\begin{eqnarray}
\Omega_{m}\left(a,\bx\right) & = & 
\frac{1}{1+ca^{3}}\left[1+\left(1+\frac{2}{3}f\left(a\right)\right)\delta\left(a,\bx\right)\right], 
\label{omloc} \\
\Omega_{\CR}\left(a,\bx\right) & = & 
- \left[{1\over1+c a^{3}}+\frac{2}{3}f\left(a\right)\right] \delta\left(a,\bx\right), 
\label{oRloc} \\
\Omega_{\Lambda}\left(a,\bx\right) & = &\frac{ca^{3}}{\left(1+ca^{3}\right)}\left[1+\frac{2}{3}f\left(a\right)\delta\left(a,\bx\right)\right], 
\label{oLloc} \\
\Omega_{\cal Q}\left(a,\bx\right) & = & 0
\label{oQloc},
\end{eqnarray}
demonstrating that the importance of curvature effects grows proportional to 
the formation of structures. 
A remarkable property is that 
$\sum \Omega_i(a,{\bf x}) = 1$ holds not only for the FLRW background, but also
at the level of perturbations. For linear perturbations the kinematic 
backreaction term does not play any role, but becomes important as soon
as quadratic terms are considered.

Let us now compare these local quantities with the domain-averaged expansion 
rate and the spatial curvature \citep{li:scale}. 
From the definition of the average $\average{}$ in (\ref{eq:Def-Mittel}) we find 
that, in principle,
fluctuations in the volume element $d\mu_{g}$ have to be taken into account. 
Writing ${\rm d}\mu_{g} = J \rmd^{3}x$ with the functional determinant 
$J=a^{3}\left(1-3\psi^{\left(1\right)}\right)$, the average over the perturbed
hypersurface agrees with an average over an unperturbed Euclidean domain 
\begin{equation}
\average{O^{\left(1\right)}}=\frac{\int_{\cD}O^{\left(1\right)}J\mbox{d}{\bf x}}{\int_{\cD}J\mbox{d}{\bf x}}%
\simeq\frac{\int_{\cD}O^{\left(1\right)}\mbox{d}{\bf x}}{\int_{\cD}\mbox{d}{\bf x}}=:\av{O^{\left(1\right)}},
\end{equation}
if we restrict our attention to linear perturbations. 

We express domain-averaged quantities in terms of the volume scale factor $a_{\cal D}$,
because we assume that the measured redshift in an inhomogeneous universe is 
related to the average scale factor. This has been advocated by \cite{rasanen:light},
where the relation
\begin{equation}
\left(1+z\right)\approx a_{\cD}^{-1}
\label{eq:aDofz}
\end{equation}
has been established. Note that in principle one would have to introduce averaging 
on some larger scale than $\cD$ to connect this background average on some domain 
$\CB$ to the redshift. For the sake of simplicity, and because we content ourselves with 
small redshifts, we use the same domain $\cD$. 
This limits the validity of the result to small redshifts.

In order to relate $a$ and $a_\cD$, we start from
\begin{equation}
H_{\cal D} = \frac{1}{3}\average{\theta} 
= \frac{\dot{a}_{\cD}}{a_{\cD}} 
= \frac{1}{a} \frac{a'_{\cD}}{a_{\cD}} = \frac{1}{a}\left(\frac{a'}{a}-\av{{\psi^{\left(1\right)}}'}\right).
\end{equation}
To first order this relation gives
\begin{equation}
a_{\cD} = a 
\left(1-\frac{1}{3}\left(\average{\delta\left(a\right)}-\average{\delta\left(1\right)}\right)\right)\;.
\label{eq:aDofa}
\end{equation}

We finally obtain the averaged Hubble rate 
\begin{equation}
H_{\cD} = H_{0} \sqrt{\frac{\Omega_{m}}{a_{\cD}^{3}}} \sqrt{1 + c a_\cD^3} 
\left[1- \frac{5\frac{a_{\cD}}{D(a_\cD)} - 3 \frac{D_0}{D(a_\cD)}}{6(1 + c a_\cD^3)}
\frac{D(a_\cD)}{D_{0}} \average{\delta_{0}} \right]
\end{equation}
and the averaged spatial curvature
\begin{equation}
\average{\CR} = 10\, \Omega_{m}\frac{H_{\text{0}}^{2}}{a_{\cD}^{2}}\frac{\average{\delta_{0}}}{D_{0}}.
\end{equation}
For later convenience we also define the function
\begin{equation} 
f_\cD \left(a_{\cD}\right) :=
\frac{5\frac{a_{\cD}}{D_{0}}-3}{2(1 + c a_\cD^3)},
\label{eq:fofaD}
\end{equation}
which is our modified version of the growth rate of Eq.~(\ref{eq:growthrate}), 
multiplied by $D\left(a\right)/D_{0}$. It basically encodes the
deviation of the time evolution of the Hubble perturbation from the time
evolution of the ensemble averaged Hubble rate
\begin{equation}
\overline{H_{\cD}}\left(a_{\cD}\right)=H_{0}\sqrt{\frac{\Omega_{m}}{a_{\cD}^{3}}} \sqrt{1 + c a_\cD^3},
\label{eq:HDofaD}
\end{equation}
as may be seen from the resulting expression
\begin{equation}
H_{\cD}=\overline{H_{\cD}}\left(a_{\cD}\right)\left(1-\frac{1}{3}f_\cD\left(a_{\cD}\right)\average{\delta_{0}}\right)\;.
\end{equation}

In the Einstein-de Sitter limit ($c\rightarrow0$ and $\Omega_{m} \rightarrow1$)
we arrive at 
\begin{equation}
H_{\cD} = \frac{H_{0}}{a_{\cD}^{3/2}}
\left(1- \frac{1}{3} a_{\cD} \average{\delta_{0}}\right),
\end{equation}
\begin{equation}
\average{\CR} = 10\frac{H_{\text{0}}^{2}}{a_{\cD}^{2}}
\average{\delta_{0}}.
\end{equation}
In order to compare this with the results of \cite{li:onset,li:scale}, we define
the peculiar gravitational potential $\varphi\left({\bf x}\right)$ via
\begin{equation}
\Delta\varphi\left(\bx\right) \equiv 4\pi G\rho^{\left(1\right)}a^{2} =
\frac{3}{2}H_{0}^{2}\frac{\delta}{a}=\frac{2}{3}\frac{1}{t_{0}^{2}}\frac{\delta}{a} 
\end{equation}
and obtain
\begin{equation}
H_{\cD}=\frac{2}{3t_{0}}a_{\cD}^{-3/2}\left[1-\frac{1}{2}a_{\cD}t_{0}^{2}\av{\Delta\varphi}\right]
\label{eq:hubEdS}
\end{equation}
and
\begin{equation}
\average{\CR} = \frac{20}{3}a_{\cD}^{-2}\av{\Delta\varphi}.
\end{equation}
While our results agree for the spatial curvature, $H_{\cD}$ is different from the result in 
\cite{li:scale}, because there the assumption $a\ll 1$ was made when applying 
(\ref{eq:aDofa}). 

Let us now turn to the dimensionless $\Omega^\cD$-parameters.
To first order, they may be expressed as
\begin{eqnarray}
\Omega_{m}^{\cD}\left(a_{\cD}\right)&=&
{1\over 1+c\, a_{\cD}^{3}} 
\left[1+\left(1+\frac{2}{3}f_\cD\left(a_{\cD}\right)\right)\,\average{\delta_{0}}\right],
\label{eq:omMofaD}\\
\Omega_{\CR}^{\cD}\left(a_{\cD}\right)&=&
- \left[{1 \over 1+c\,a_{\cD}^{3}} + \frac{2}{3}f_\cD\left(a_{\cD}\right)\right]\,\average{\delta_{0}},
\label{eq:dummy-3}\\
\Omega_{\Lambda}^{\cD}\left(a_{\cD}\right)&=&
{c\,a_{\cD}^{3}\over 1 + c a_\cD^3} \left[1+\frac{2}{3}f_\cD\left(a_{\cD}\right)\,
\average{\delta_{0}}\right],\\
\Omega_{{\cal Q}}^{\cD}\left(a_{\cD}\right)&=&0.
\label{eq:omQofaD}
\end{eqnarray}
When taking the limit $\cD\rightarrow0$ in 
Eqs.~(\ref{eq:omMofaD}) -- (\ref{eq:omQofaD}), we recover the point-wise defined $\Omega$-parameters of Eqs.~(\ref{omloc}) -- (\ref{oQloc}). This provides a self-consistency check of 
the averaging framework.

From the expressions for the $\Omega^\cD$-parameters one can easily calculate the ensemble averages
and the ensemble variance. $\overline{\average{\delta_{0}}}=0$,
since the domain-averaged overdensity of $\cD$, in general non-zero, 
averages out when we consider a large number of domains of given size 
and local density fluctuations drawn from the same (Gaussian) distribution. 

Here we adopt the common view that linear theory is a good description of the 
present universe at the largest observable scales (which has been questioned recently
in \citealt{rasanen:pert}). 
We then find the ensemble average of the
curvature parameter $\overline{\Omega_{\CR}^{\cD}}$ to vanish.
For the matter density parameter Eq.~(\ref{eq:omMofaD}) yields 
\begin{equation}
\overline{\Omega_{m}^{\cD}}\left(a_{\cD}\right)=\left(1+c\,a_{\cD}^{3}\right)^{-1}\;.
\label{eq:Omegambar}
\end{equation}
This may be used to verify that the relation $\overline{\Omega_{m}^{\cD}}+\overline{\Omega_{\Lambda}^{\cD}}=1$
holds. In addition, this relation implies that 
$\overline{\Omega_{m}^{\cD}}\left(a_{\cD_{0}}\right)$ corresponds to 
today's background matter density parameter:
\begin{equation}
\overline{\Omega_{m}^{\cD}}\left(a_{\cD_{0}}\right)=\Omega_{m} + 
\CO\left(\overline{\average{\delta_0^2}}\right)\;.
\end{equation}

However, this is true at first order in the density contrast only, because in this 
case ensemble averages agree with background quantities. 
At higher orders, the ensemble averages differ from the background quantities.

\section{Variances of locally averaged cosmological parameters}
\label{sec:Var-loc-cosm-par}

After having convinced ourself that the expectations of the averaged 
$\Omega^\cD$-parameters are identical to their $\Lambda$CDM background 
values up to second--order corrections, we now turn to the study of their ensemble 
variances.

All variances of domain averaged cosmological parameters can be related 
to the variance of the overdensity of the matter distribution, 
$\sigma\left(\average{\delta_{0}}\right)$. 

In order to specify $\average{\delta_{0}}$, we introduce the normalized window function 
$W_{\cD}\left(X\right)$ and write 
\begin{eqnarray}
\average{\delta_{0}} & = & \int_{\mathbb{R}^{3}}\delta_{0}\left(\bx\right)W_{\cD}\left(\bx\right)\rmd^{3}x\nonumber \\
 & = & \int_{\BR^{3}}\widetilde{\delta}_{0}\left(\bk\right)\widetilde{W}_{\cD}\left(\bk\right) \rmd^{3}k,
\label{eq:dummy-2}
\end{eqnarray}
A tilde denotes a Fourier--transformed quantity. 
With the definition of the matter power spectrum
\begin{equation} 
\overline{\widetilde{\delta}_{0}\left(\bk\right)
\widetilde{\delta}_{0}\left(\bk'\right)} =
\delta^{\rm Dirac}\left(\bk+\bk'\right)\ P_{0}\left(k\right),
\end{equation} 
where $\delta^{\rm Dirac}$ denotes Dirac's delta function,
the ensemble variance of the matter overdensity becomes
\begin{equation}
\left(\sigma_{\cD_{0}}\right)^{2}:=\sigma^{2}\left(\average{\delta_{0}}\right)=\int_{\BR^{3}}P_{0}\left(k\right)\widetilde{W}_{\cD}\left(\bk\right)\widetilde{W}_{\cD}\left(-\bk\right)\rmd^{3}k.
\label{eq:sigma}
\end{equation}

For a spherical window function, this expression is the well--known matter variation in a sphere, often used to normalize
the matter power spectrum by fixing its value for a sphere with a
radius of $8 h^{-1} \mathrm{Mpc}$ ($\sigma_{8}$). To calculate
this variance, we assume a standard $\Lambda$CDM power spectrum in the parametrization
of \cite{eisenst-hu}.
Knowing $\sigma_\cD$
at a particular epoch of interest, we can calculate all the fluctuations
in the cosmic parameters. They read:
\begin{eqnarray}
\delta H_{\cD}&=&\frac{1}{3}\overline{H_{\cD}}\left(a_{\cD}\right)f_\cD
\left(a_{\cD}\right)\sigma_{\cD_{0}}\;,%
\label{eq:sigH}\\
\delta\Omega_{m}^{\cD}&=&\overline{\Omega_{m}^{\cD}}\left(a_{\cD}\right)
\left(1+\frac{2}{3}f_\cD\left(a_{\cD}\right)\right)\sigma_{\cD_{0}}\;,%
\label{eq:sigOm}\\
\delta\Omega_{\CR}^{\cD}&=&\overline{\Omega_{m}^{\cD}}\left(a_{\cD}\right)
\left(1+\frac{2}{3}\frac{f_\cD \left(a_{\cD}\right)}{\overline{\Omega_{m}^{\cD}}
\left(a_{\cD}\right)}\right)\sigma_{\cD_{0}}\;,\label{eq:sigOR}\\
\delta\Omega_{\Lambda}^{\cD}&=&\overline{\Omega_{\Lambda}^{\cD}}\left(a_{\cD}\right)
\frac{2}{3}f_\cD \left(a_{\cD}\right)\sigma_{\cD_{0}}\;,\label{eq:sigOL}\\
\delta\Omega_{\CQ}^{\cD}&=&\CO\left(\left(\sigma_{\cD_{0}}\right)^{2}\right)\;, 
\label{eq:sigOQ}
\end{eqnarray}
where e.g. $\delta\Omega_{\Lambda}^{\cD}$ denotes the square root of the 
variance, $\delta\Omega_{\Lambda}^{\cD}:=\sigma\left(\Omega_{\Lambda}^{\cD}\right)$.
$\overline{\Omega_{m}^{\cD}}$, $\overline{H_{\cD}}$ and 
$f_\cD\left(a_{\cD}\right)$ were defined in Eq.~(\ref{eq:Omegambar}), 
(\ref{eq:HDofaD}) and (\ref{eq:fofaD}) and 
$\overline{\Omega_{\Lambda}^{\cD}}=1-\overline{\Omega_{m}^{\cD}}$.

These variances are the minimum ones that one can hope to obtain
by measurements of regions of the universe of size $\cD$. They
do not include any observational uncertainties, nor biasing or sampling
issues. They are intrinsic to the inhomogeneous dark matter 
distribution that governs the evolution of the Universe. 

\begin{figure}
\includegraphics[width=\linewidth]{%
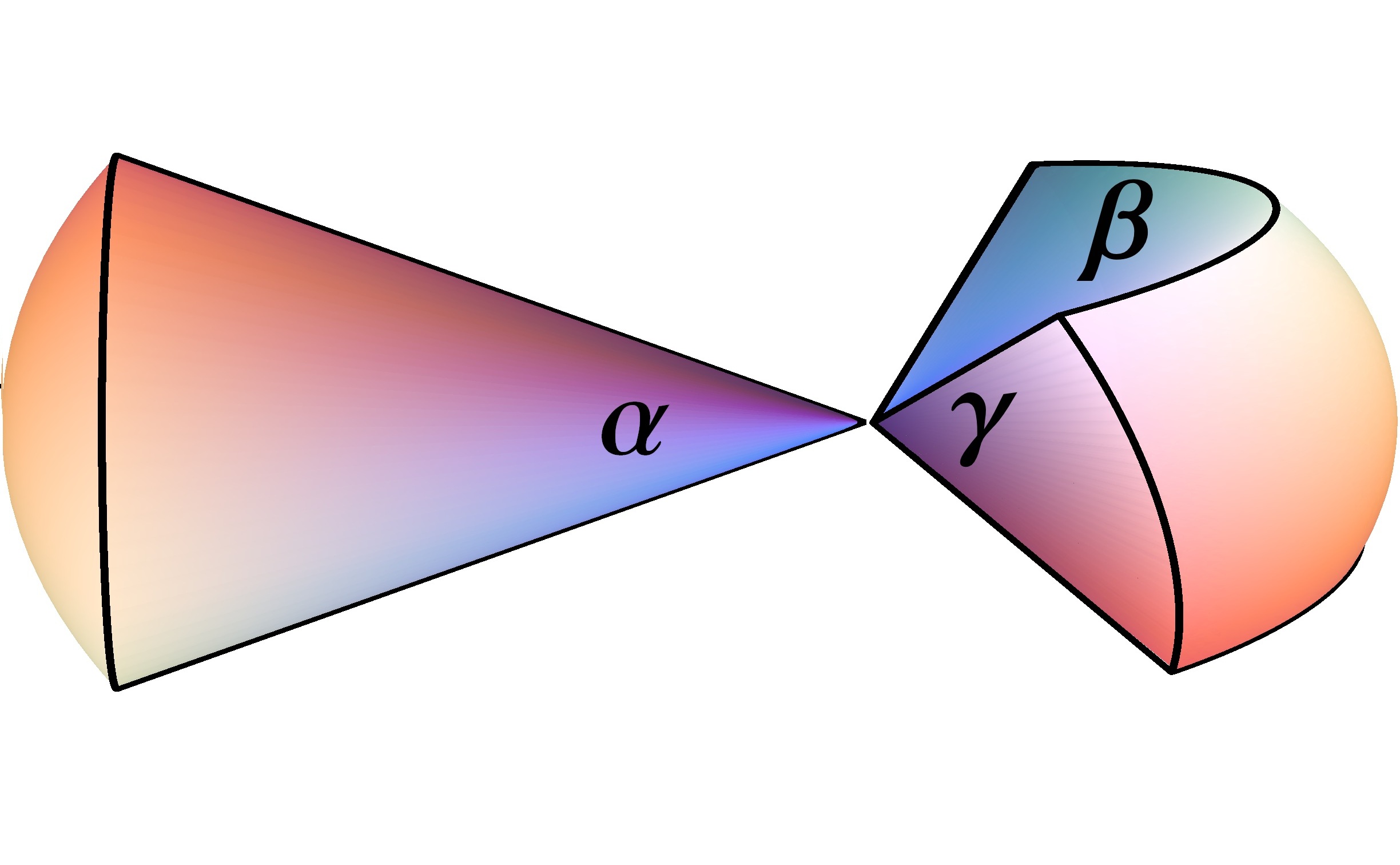}%

\caption{
The two survey geometries considered (separately). A simple
cone with one single opening angle $\alpha$ and a slice given by
two angles $\beta$ and $\gamma$. 
\label{fig:geometry}}
\end{figure}

Equations (\ref{eq:sigH}) to (\ref{eq:sigOQ}) are interesting in two respects: 
Firstly, our expression for $\delta H_{\cD}$ is simpler than the one 
in \cite{clarkson:hubble}, nevertheless, both results agree. 
Secondly, Eqs.~(\ref{eq:sigH}) to (\ref{eq:sigOQ}) quantify
the connection between fluctuations in cosmological parameters and
inhomogeneities in the distribution of matter.
If we choose "today" as our reference value, Eqs.~(\ref{eq:sigH}) to 
(\ref{eq:sigOL}) allow us to predict the domain averaged cosmological
parameters: 
\begin{equation}
\begin{array}{ccllcl}
H_{\cD} & = & H_{0}&\pm&\frac{1}{3}H_{0}\, f_{\cD_0}&\sigma_{\cD_{0}}%
\\
\Omega_{m}^{\cD} & = & \Omega_{m}&\pm&\Omega_{m}\left(1+\frac{2}{3}f_{\cD_0}\right)&\sigma_{\cD_{0}}%
\\
\Omega_{\CR}^{\cD} & = & 0&\pm&\left(\Omega_{m}+\frac{2}{3}f_{\cD_0}\right)&\sigma_{\cD_{0}}\,%
\\
\Omega_{\Lambda}^{\cD} & = & \Omega_{\Lambda}&\pm&\frac{2}{3}\Omega_{\Lambda}\,f_{\cD_0}&\sigma_{\cD_{0}}%
\end{array}
\label{eq:central}
\end{equation}
with
\begin{equation}
f_{\cD_0} \equiv f_\cD \left(a_{\cD_{0}}\right)
=\frac{\Omega_{m}}{2}\left(5 D_{0}^{-1}-3\right)\approx\left\lbrace\begin{array}{ll}0.5
& \Lambda\mathrm{CDM}\\1.0 & \mathrm{EdS}\end{array}\right. ,
\end{equation}
where we assumed $\Omega_{m}=0.3$ for $\Lambda\mathrm{CDM}$. More generally,
for  $\Omega_{m}>0.1$, $f_{\cD_0}$ may be approximated by \citep{lahav:fit,eisenst-hu}
\begin{equation}
f_{\cD_0}\approx\frac{1}{140}\left(2+140\,\Omega_{m}^{4/7}-\Omega_{m}-\Omega_{m}^{2}\right). 
\end{equation}

From the knowledge of $\sigma_{\cD_{0}}$
we may therefore easily derive the variation of cosmological
parameters.
To relate our calculations to real surveys, we elaborate in the next section on how to calculate
$\sigma_{\cD_{0}}$ for several survey geometries.

\section{The effect of the survey geometry
\label{sec:survey-geom}}

\begin{figure}
\includegraphics[width=\linewidth]{%
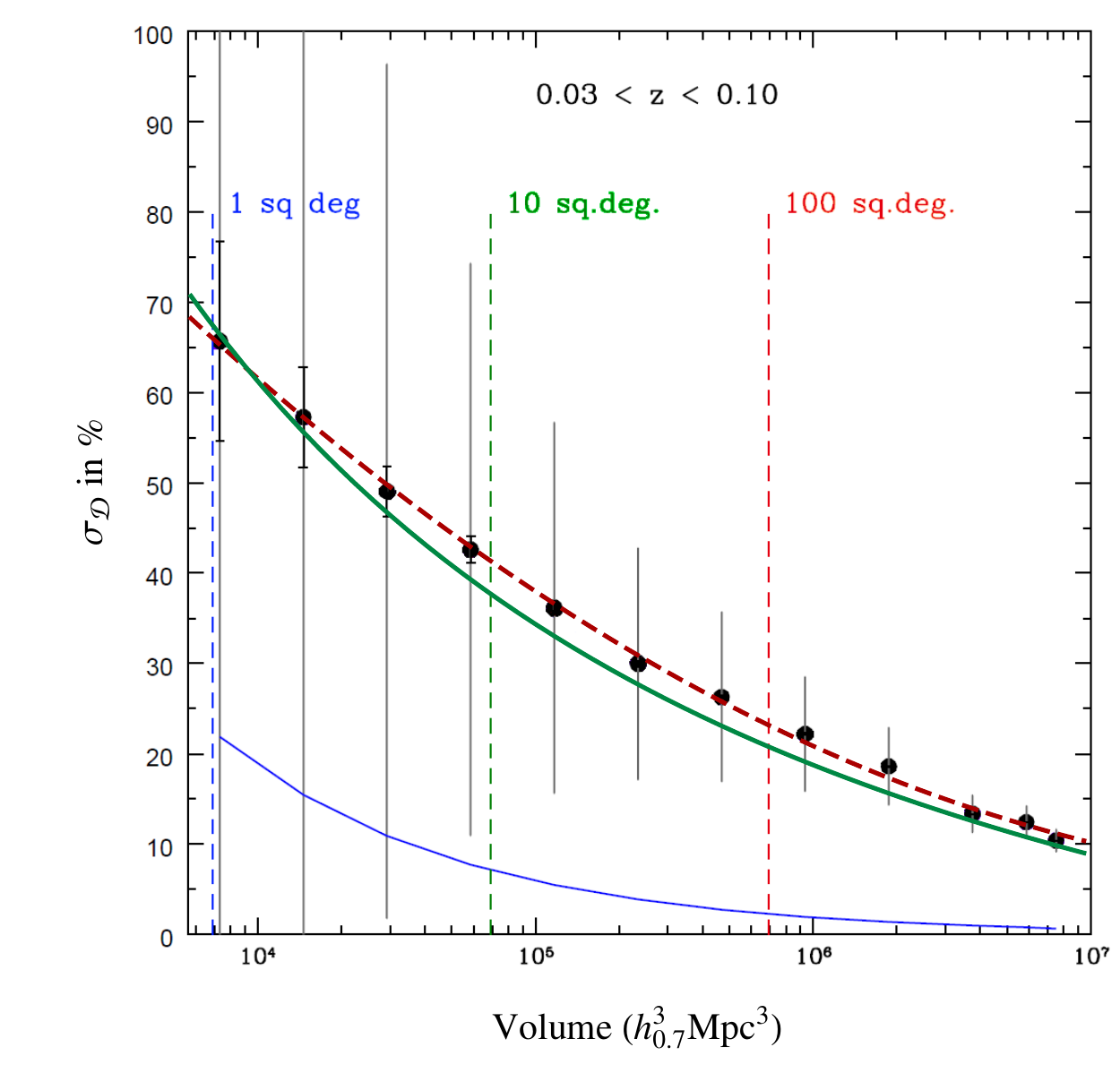}%
\caption{
Variance of the matter density, $\sigma_\cD$, as a function of the 
observed domain volume. Data were derived from the SDSS main sample by 
\cite{driver:cosvar}. The dashed (red) line shows the fit of \cite{driver:cosvar} to the data,
the solid (green) line is our result including the sample variance [solid (blue) line at the bottom].
\label{fig:driver}}
\end{figure}

Observations of the universe are rarely full--sky measurements and
typically sample domains much smaller than the Hubble volume.
We therefore must address the problem of the survey geometry.
Effects from a limited survey size are in particular important for deep fields, 
as studied for example in \cite{moster:cookbook}. While in their case, for small angles and 
deep surveys, approximating the observed volume by a rectangular geometry is appropriate,
it probably is not appropriate for the bigger survey volumes that we have in mind.

We therefore chose two different geometries that resemble observationally
relevant ones. Firstly, we used a simple cone with a single opening 
angle $\alpha$. The second geometry is a slice described by two angles 
$\beta$ and $\gamma$ for the size in right ascension and declination 
respectively. In the radial direction we assumed a top hat window, whose 
cut--off value corresponds to the depth of the survey. Both shapes are shown 
in Fig.~\ref{fig:geometry}.

To calculate $\sigma_{\cD_{0}}$ for both geometries, we used a decomposition 
into spherical harmonics. This allowed us to derive an expression for the expansion 
coefficients in terms of a series in $\cos\left(2n\alpha\right)$ for the 
cone and a similar one for the slice, depending on trigonometric 
functions of $\beta$ and $\gamma$. The radial coefficients were calculated
numerically using the $\Lambda$CDM power spectrum of 
\citet{eisenst-hu}, including the effect of baryons on the overall shape 
and amplitude of the matter power spectrum, but without baryon 
acoustic oscillations. 

All plots use best-fit $\Lambda$CDM values as given in 
\cite{wmap-7}; $\Omega_{b}=0.0456$, $\Omega_{cdm}=0.227$ and $n_{s}=0.963$.
The power spectrum is normalized to $\sigma_{8}=0.809$.

To ensure that the result of our calculation for the slice-like geometry and a standard 
$\Lambda$CDM power spectrum is reasonable, we compared 
it with an analysis of SDSS data by \cite{driver:cosvar}.
In Fig.~\ref{fig:driver} we show this comparison of
their r.m.s.~matter overdensity $\sigma_\cD$ obtained from the SDSS main
galaxy sample, in terms of its
angular extension (and hence the volume). The dashed line going through the points shows their
empirical fit to the data. The solid green line shows
our result for the cosmic variance of a slice with
respective angular extension (for $\beta=\gamma$), plus their sample variance.
Note that our result is not a fit to SDSS data, but is a prediction based on 
the WMAP 7yr data analysis. Additionally, the real SDSS window function is slightly 
more complicated than our simplistic window, thus perfect agreement is not to be 
expected.

\begin{figure}
\includegraphics[width=\linewidth]{%
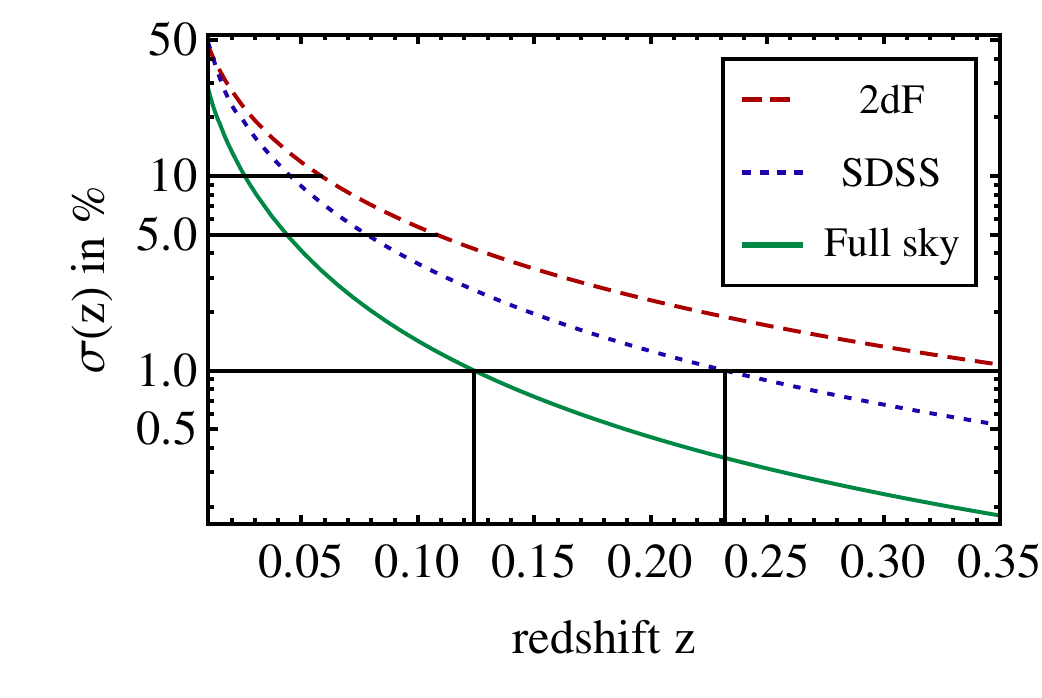}

\caption{
Variance of the matter density, $\sigma_\cD$, for survey geometries 
resembling the 2dFGRS, the SDSS and a hypothetical full sky survey as a function of 
maximum redshift considered. We find that the determination of the local $\sigma_\cD$ below
redshifts of $0.1$ (corresponding to $\sim 400$ Mpc) is fundamentally limited by 
cosmic variance to the $1\%$ level.
\label{fig:SigmaVergleich}}

\end{figure}

For the full SDSS volume, $\sigma_{\cD_{0}}$ is shown in Fig.~\ref{fig:SigmaVergleich}.
For comparison we also added the smaller, southern hemisphere 2dF survey and a 
hypothetical full sky survey. For the two surveys, we assumed an approximate angular 
extension of 120\textdegree{}$\times$60\textdegree{} for SDSS and for the two
fields of the 2dF survey 80\textdegree{}$\times$15\textdegree{} and
75\textdegree{}$\times$10\textdegree{}.
The ongoing BOSS survey corresponds to the plot for the SDSS geometry because it will basically have 
the same angular extension. Because it will target higher redshifts, it is
not in the range of our calculation, however.
As a rough statement (the precise value depends on the redshift), one
may say that the 2dF survey is a factor of 5 and the SDSS survey
a factor of 2.5 above the variance of a full
sky survey. This is interesting because the SDSS survey covers approximately
only $1/6$ of the full sky and the 2dF survey only $1/20$.
This is due to the angular dependence of $\sigma_\cD$. We find that fluctuations drop 
quickly as we increase small angles and flattens at large angles.

From Fig.~(\ref{fig:SigmaVergleich}) we see that the cosmic variance of the 
matter density for the SDSS geometry is 5\% at a depth of $z\approx0.08$ and still 1\% out to 
$z\approx0.23$. Note, however, that the extension of the domain of (spatial) averaging to a redshift 
of $0.35$ is clearly not very realistic because lightcone effects will become relevant with
increasing extension of the domain. The assumption that this domain
would be representative for a part of the hypersurface of constant
cosmic time becomes questionable. We expect, however, that in this range
evolution effects will only be a minor correction to the result presented here because of the following estimation:

The main effect we miss by approximating the lightcone by a fixed spatial hypersurface is evolution in the matter density distribution. To specify what "region that is not too extended spatially" means, one should therefore estimate the maximum evolution in a given sample. This can be done by determining the growth of the density contrast in the outermost (and therefore oldest) regions of the sample. In the linear regime considered here, the evolution of the density contrast is given by $\delta\left(z\right)=\delta_{0}D\left(0\right)/D\left(z\right)$. Therefore, lightcone corrections should be smaller than $\epsilon_{l}=1-D\left(z\right)/D\left(0\right)$ which is $\epsilon_{l}\approx5\%$ for $z=0.1$ and $\epsilon_{l}\approx14\%$ for $z=0.3$. Therefore, the order of magnitude of our results should be correct on a wider range of scales, but on scales above $z=0.3$ the corrections to our calculation are expected to pass beyond $15\%$.

Finally, it should be noted that for large volumes the actual shape of the survey geometry is not very important. As long as all dimensions are bigger than the scale of the turnover of the power spectrum, the deviation of the cosmic variance for our shapes, compared to those of a box of 
equal volume, is at the percent level.
To reach this result, we compared $\sigma_{\cD_{0}}$
for the slice--like geometry to its value for a rectangular box of the same volume.
We used a slice for which $\beta=\gamma$. The box was constructed
to have a quadratic basis and the same depth as the slice in radial direction.
Therefore the base square of the box is smaller than the square given by the two angles of the slice.
The result of this comparison is that the deviation of $\sigma_{\cD_{0}}^{rect}$
from the value for the slice is at 
most 6\% for angles above $\beta\approx10\text{\textdegree}$. For smaller angles the deviation becomes bigger
and redshift--dependent. This is caused by the changing shape of the power spectrum at small
scales. 
The large angle behavior confirms an observation of \cite{driver:cosvar}. 
They found that the cosmic variance in the SDSS dataset was the same
for both of the two geometries they considered.

\section{Fluctuations of the curvature parameter
\label{sec:curv-fluct}}

After the general study of the effect of the shape of the observational
domain $\cD$ on $\sigma_{\cD_{0}}$, one may ask for which
parameter the fluctuations are most important.

\begin{figure}
\includegraphics[width=0.96\linewidth]{%
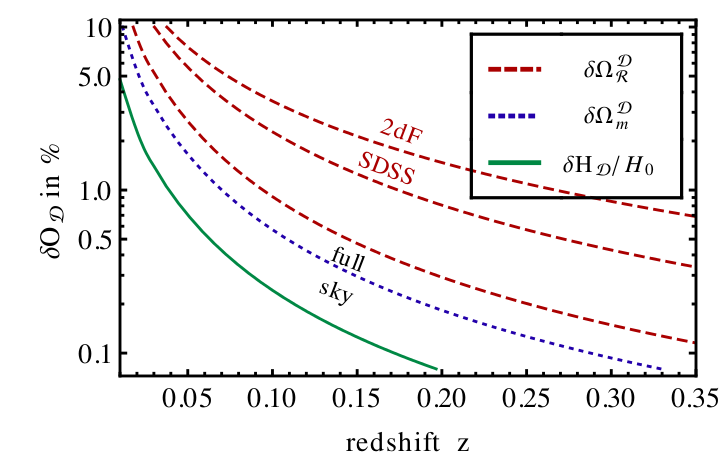}
\caption{
Top three lines: the expected r.m.s.~fluctuation of the 
curvature parameter, $\delta\Omega_{\CR}^{\cD}$, for geometries resembling the 
2dFGRS, the SDSS, and a full sky. The two lowest lines are 
the expected r.m.s.~fluctuations of the parameters $\Omega_{m}^{\cD_{0}}$ and 
$H_{\cD_{0}}$ for a full--sky survey extending to the respective redshift. The 
curvature fluctuations turn out to be higher than all other fluctuations.
\label{fig:cosvar}}
\end{figure}

The three lowest lines in the plot of Fig.~\ref{fig:cosvar} show that this is the case for the
curvature fluctuations.
The two lowest lines, showing the fluctuations $\delta\Omega_{m}$
and $\delta H_{\cD_{0}}/H_{\cD_{0}}$ for the full sphere, lie a factor
of 1.6 and 3.8, respectively, below the respective curvature fluctuations $\delta\Omega_{\CR}^{\cD}$. Therefore the
fluctuations of $\Omega_{m}$ play a smaller role for all universes with $\Omega_{m}<1$.
The uncertainty
in $H_{\cD_{0}}$, which has been in the focus of the investigations
so far \citep{hubble:shiturner,li:scale,clarkson:hubble}, contributes
even less to the distortion of the geometry, as we shall discuss
in Section~\ref{sec:Variances-on-BAO}.

What this means for real surveys, such as the 2dF or the SDSS survey, is
shown by the three upper lines in Fig.~\ref{fig:cosvar}. They compare
$\delta\Omega_{\CR}^{\cD}$ for the slices observed by these surveys
to that of a full sky measurement.
$\delta\Omega_{\CR}^{\cD}$
is bigger than one percent up to a redshift of $0.18$ for the SDSS
and $0.28$ for the 2dF survey and it does not drop below $0.001$
for values of $z$ as high as $0.5$. This may seem very low, but
it has been shown that getting the curvature of the universe wrong
by $1\permil$ already affects our ability to measure the dark energy
equation of state $w\left(z\right)$ \citep{clarkson:curvature}. Of
course one has to keep in mind that for high redshifts one has to
be careful with the values presented here because they are based on the
assumption that the observed region lies on one single spatial hypersurface.
Because this approximation worsenes beyond a redshift of $0.1$,
there may be additional corrections to the size of the fluctuations stemming from
lightcone effects.

\begin{figure*}
\includegraphics[width=0.5\textwidth]{%
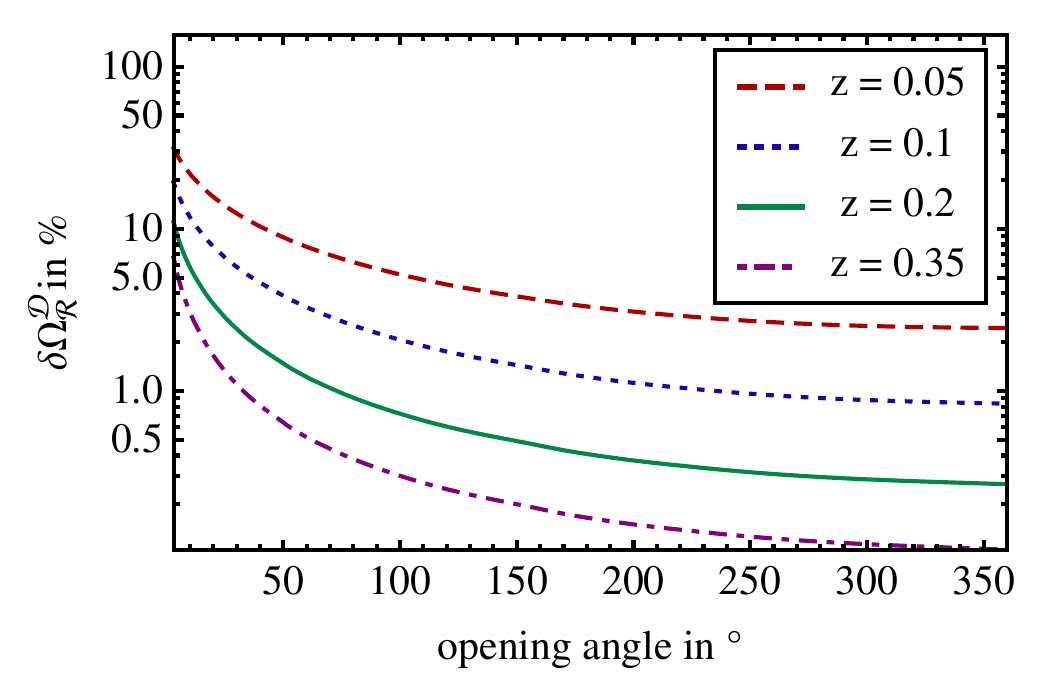}%
\includegraphics[width=0.5\textwidth]{%
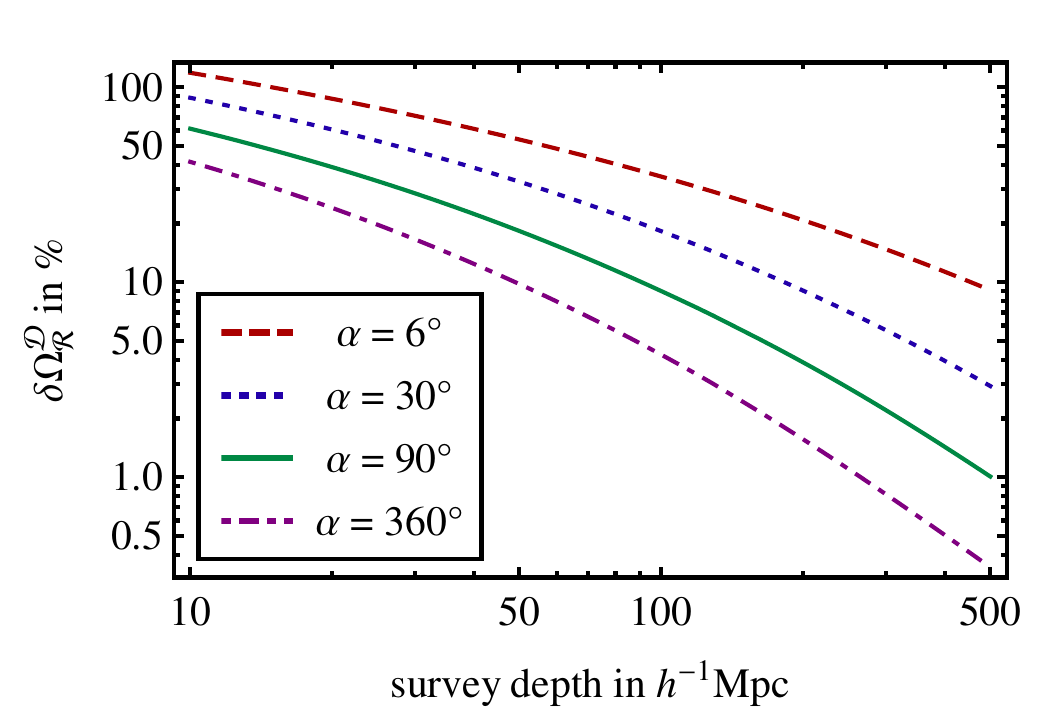}
\caption{
Cosmic variance of the curvature parameter.
{\it Left panel:} Dependence of $\delta\Omega_{\CR}^{\cD}$ on the opening angle of 
the cone-like survey geometry of Fig.~\ref{fig:geometry}
for different top hat depths of the survey. \label{fig:winkdep}
{\it Right panel:} Dependence of $\delta\Omega_{\CR}^{\cD}$
on the depth of the survey. For a small cone of 6\textdegree{} opening
angle we expect curvature fluctuations of 10\% up to $450 h^{-1}\mathrm{Mpc}$.
\label{fig:raddep}
}
\end{figure*}

To investigate the curvature fluctuations for more general geometries, 
we show in Fig.~\ref{fig:winkdep} the angular and radial dependence 
of the curvature fluctuation $\delta\Omega_{\CR}^{\cD}\left(\alpha\right)$ 
for the cone-like window of Fig.~\ref{fig:geometry}.

On the l.h.s. of Fig.~\ref{fig:winkdep} we evaluate the angular dependence.
For a survey that only reaches a redshift of $0.1$, the fluctuations
are still higher than $0.01$ for a half--sky survey. It is interesting
to note that for a deeper survey, $\delta\Omega_{\CR}^{\cD}\left(\alpha\right)$
grows much faster when $\alpha$ is reduced than for a shallow survey.
This is because $\sigma_{\cD_{0}}\left(R\right)$ changes from a
relatively weak $R^{-1}$ decay to a $R^{-2}$ decay on larger scales.
For $z=0.35$, this behavior dominates and a decrease
in $\alpha$ increases $\sigma_{\cD_{0}}\left(R,\alpha\right)$ stronger
than in the $R^{-1}$ regime.

On the r.h.s. of Fig.~\ref{fig:raddep} we show the dependence
of $\delta\Omega_{\CR}^{\cD}$ on the survey depth for some opening
angles of the cone-like window. For narrow windows the fluctuation
in $\Omega_{\CR}^{\cD}$ stays high, even beyond the expected homogeneity
scale of $100 h^{-1} \mathrm{Mpc}$. For $R=200 h^{-1} \mathrm{Mpc}$ and a 
$6\text{\textdegree}$
window, for example, it is still at $\delta\Omega_{\CR}^{\cD}\approx0.2$.
For smaller beams these fluctuations persist even out to much longer
distances. Therefore they play an important role for deep field galaxy surveys,
as shown in \cite{moster:cookbook,driver:cosvar} for the matter density
fluctuations. But even for wider angles, fluctuations in curvature
persist on sizeable domains. If one recalls that the distance given
for the full sphere of 360\textdegree{} is its radius, this means
that regions in the Universe as big as $540 h^{-1} \mathrm{Mpc}$ have
typical curvature fluctuations on the order of 1\%. This is not
that small because the last scattering surface at $z\approx1100$ is only
$9600 h^{-1} \mathrm{Mpc}$ away. One of these regions therefore fills more than
5\% of the way to that surface.

\begin{figure}
\includegraphics[width=\linewidth]{%
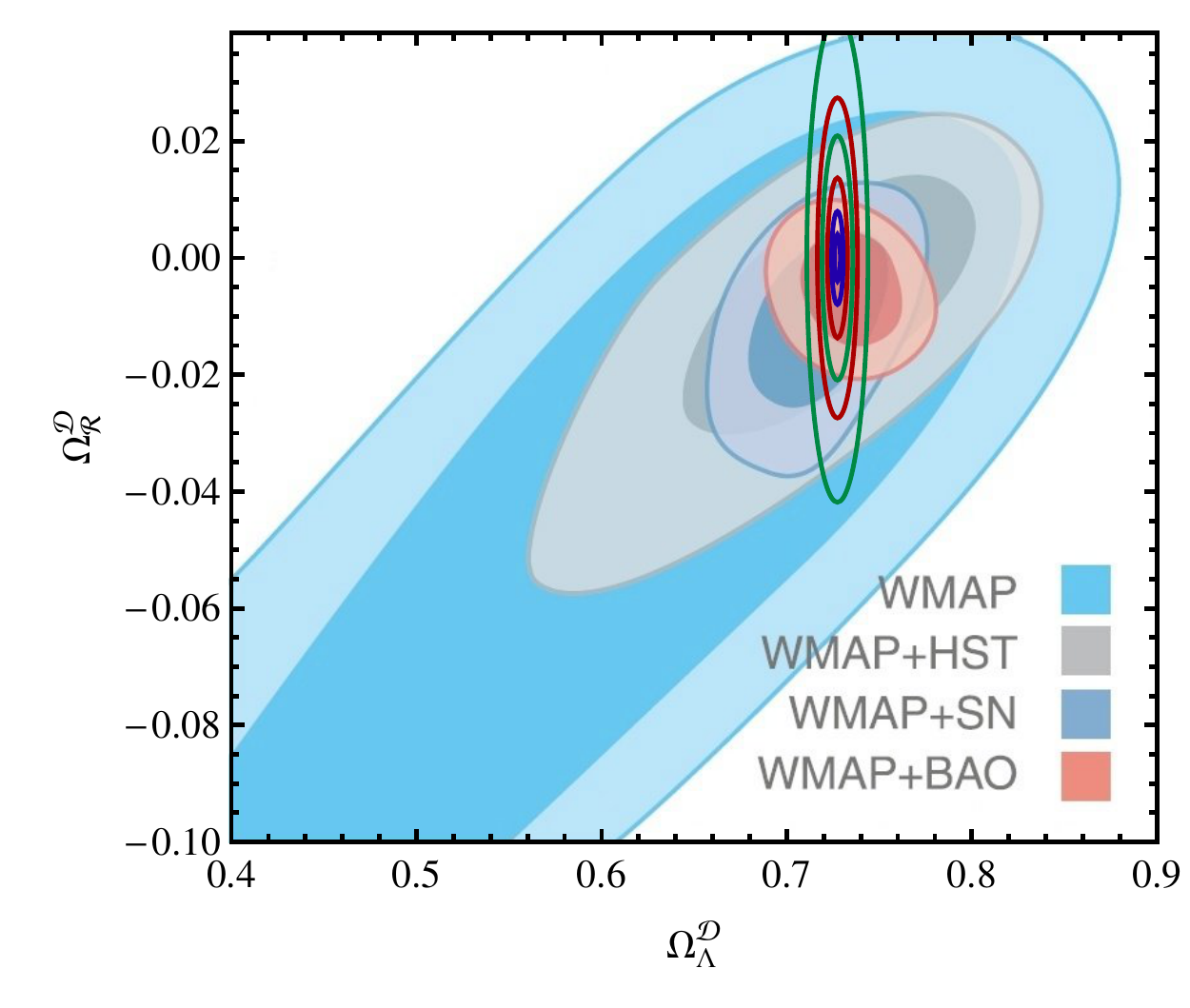}
\caption{
{
Minimum confidence contours in $\Omega_{\Lambda}^\cD$ and $\Omega_{\CR}^{\cD}$
achievable in different volumes through fluctuations of matter. 
The green (outermost) ellipses are the $95\%$ and $68\%$ contours for the volume from 
which the HST data are drawn. The next inner (red) ones
are for a survey of the size of the 2dF survey up to $z=0.2$. In the middle there is
a small double ellipse in blue, showing the values for the SDSS volume up to $z=0.3$.
The background image depicts the results from
WMAP~5 \citep{wmap}. They give the experimental values and uncertainties 
on these parameters for a combination of various experimental probes.}
\label{fig:WMAP}}
\end{figure}

To put these values into perspective, we compare the WMAP~5yr confidence
contours \citep{wmap} on the curvature parameter with those that may in principle
be derived from the 2dF or the SDSS survey in Fig.~\ref{fig:WMAP}. 
Because they only sample a finite size of the
Universe, one cannot be sure that this value is indeed the background
value and not only a local fluctuation. The cosmic variance induced
by this finite size effect is, for the 2dF survey volume up to $z\approx0.2$,
shown by the two second--largest (red) 
ellipses. The two innermost (blue) ones depict the minimum
possible error using the SDSS survey volume up to $z\approx0.3$.
Clearly, the determination of $\Omega_{\CR}^{\cD}$ may
perhaps be improved by a factor of two if one were to eliminate all
other sources of uncertainty. This may be less if lightcone effects
play a non--negligible role already for $z\approx0.3$.

\begin{figure}
\includegraphics[width=\linewidth]{%
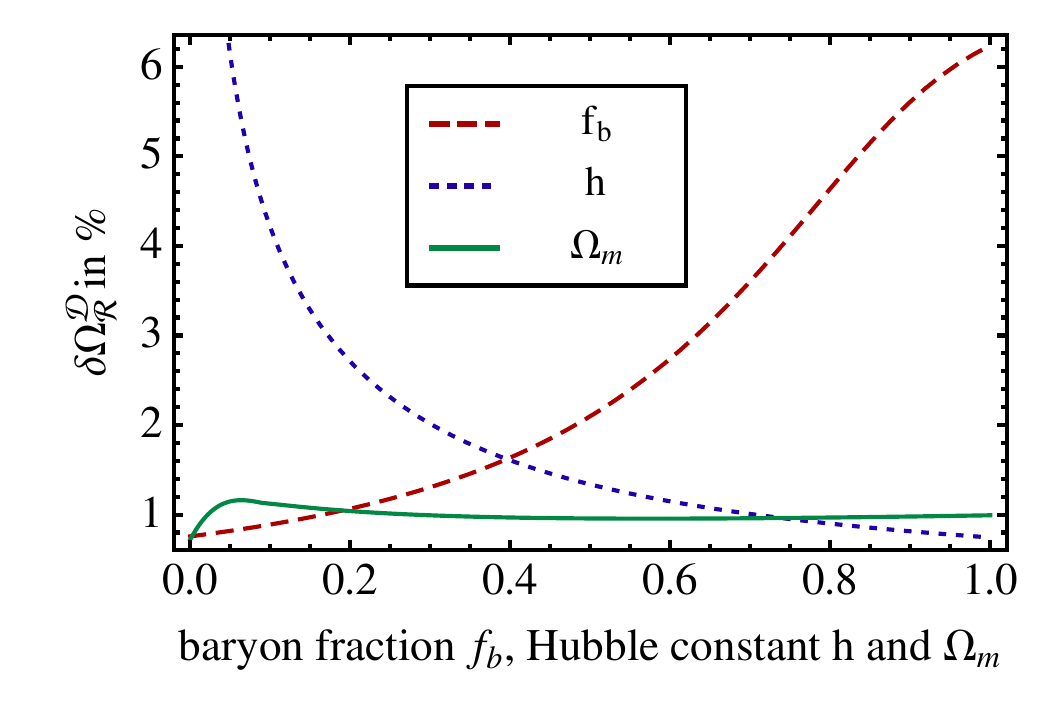}
\caption{
Dependence of $\delta\Omega_{\CR}^{\cD}$ on some cosmological
parameters for a spherical domain extending to $z=0.09$. 
The basis is the $\Lambda$CDM model with $\Omega_{b}=0.0456$,
$\Omega_{cdm}=0.227$, $h=0.7$, $n_{s}=0.963$ and $\sigma_{8}=0.809$.
For this model and for the chosen redshift, 
$\delta\Omega_{\CR}^{\cD}\approx0.01$. We then varied
$\Omega_{m} = \Omega_{b}+\Omega_{cdm}$, $f_{b}=\Omega_{b}/\Omega_{m}$
and $h$ between $0$ and $1$, holding the other parameters fixed
at their aforementioned values. Because $\sigma_{8}$ is fixed, the fluctuation
in $\Omega_{\CR}^{\cD}$ is nearly independent on $\Omega_{m}$.
\label{fig:pardepdOR}}
\end{figure}

Fig.~\ref{fig:pardepdOR} shows the dependence of the curvature
fluctuation on the considered cosmology. For this study we fixed the
spectral index and the normalization of the spectrum to $n_{s}=0.963$
and $\sigma_{8}=0.809$, respectively. We varied each of the other parameters
one after another, while keeping the remaining ones fixed to the concordance
values. We used the SDSS geometry out to a redshift of $z=0.09$ as a reference 
value at which we conducted this investigation, because the concordance values lead
to a $\delta\Omega_{\CR}^{\cD}$ of $0.01$ for this configuration. Interestingly
enough, the dependence on the $\Omega_{m}$ parameter is very weak.
This means that the value does not differ much for the flat $\Lambda$CDM model and
the EdS model. This is surprising, because the prefactor of $\sigma_{\cD_{0}}$ in (\ref{eq:sigOR})
changes by a factor of 3 from about 5/11 for $\Lambda$CDM to 5/3
for EdS. This rise, however, is compensated for by a drop of the value of $\sigma_{\cD_{0}}$.
The reason for this drop is that a higher $\Omega_{m}$ leads to more power on small scales. Because
we kept the integrated normalization fixed at a given value of $\sigma_{8}$,
this means less power on large scales, i.e.~at $z=0.09$.
Moreover, a variation of the Hubble constant $h$ and the baryon fraction
$f_{b}$ has only a small effect around the concordance value. $\delta\Omega_{\CR}^{\cD}$
changes significantly only for more extreme values of $f_{b}$ and $h$.

\section{Fluctuations of the acoustic scale
\label{sec:Variances-on-BAO}}

Let us now turn to the effect of fluctuations caused by inhomogeneities 
on the local distance estimates. An important distance measure, recently used
in BAO experiments, is $D_{V}$. It was introduced in \cite{eisenstein:BAO}
and mixes the angular diameter distance and the comoving coordinate distance to
the BAO ring. It is measured through the BAO radius perpendicular to the line
of sight $r_{\perp}$ and the comoving radius parallel to the line of sight $r_{\parallel}$. 
\begin{equation}
r_{bao}:=\left(r_{\parallel}r_{\perp}^{2}\right)^{\frac{1}{3}}=D_{V}\left(z\right)\Delta\theta^{2}\frac{\Delta z}{z}
\end{equation}
One can, therefore, determine the distance $D_{V}$ to the corresponding redshift, if the 
comoving radius of the baryon ring $r_{bao}$ is known.
This may be achieved by a measurement of the angle of the BAO ring on the sky
$\Delta\theta$ and its longitudinal extension $\Delta z/z$.
The precise definition of $D_{V}$ is derived from the expressions of the comoving
distances $r_{\parallel}$ and $r_{\perp}$:
\begin{eqnarray}
r_{\parallel} &=& 
\intop_{z}^{z+\Delta z}\frac{c}{H(z^{\prime})}{\rm d}z^{\prime}\approx\frac{c\Delta z}{H(z)}=
\frac{cz}{H(z)}\frac{\Delta z}{z}, \\
r_{\perp} &=& \left(1+z\right)D_{A}(z)\Delta\theta, 
\end{eqnarray}
from which we find
\begin{equation}
D_{V}(z)=\left(\frac{cz}{H(z)}D_{M}^{2}(z)\right)^{\frac{1}{3}}, 
\end{equation}
where $D_{M}$ is the comoving angular distance
\begin{equation}
D_{M}(z)=c\left(\sqrt{\Omega_{k}}H_{0}\right)^{-1}\sinh\left(\sqrt{\Omega_{k}}I\left(z\right)\right),
\end{equation}
with
\begin{equation}
I(z)=\intop_{0}^{z}\frac{H_{0}}{H(z^{\prime})}{\rm d}z^{\prime}.
\end{equation}

As already mentioned above, the term $\Omega_{\CQ}^{\cD}$ vanishes
in our first--order treatment and the curvature contribution
scales as $a_{\cD}^{-2}$. Therefore we may express the Hubble rate as 
\begin{equation}
\frac{H_{\cD}(z)}{H_{\cD_{0}}} = 
\left[(1+z)^3 \Omega_{m}^{\cD_{0}} + (1+z)^2 \Omega_{\CR}^{\cD_{0}} 
+ (1-\Omega_{m}^{\cD_{0}}-\Omega_{\CR}^{\cD_{0}})\right]^{\frac 12}, 
\label{eq:dummy-1}
\end{equation}
where we assumed the relation between redshift and average scale factor of 
Eq.~(\ref{eq:aDofz}). We may now calculate the fluctuation of $D_{V}$,
\begin{eqnarray}
\frac{\delta r_{\parallel}}{r_{\parallel}} & = &
 \frac{\delta H_{\cD_{0}}}{H_{\cD_{0}}} +
 \left|\frac{1-(1+z)^{2}}{2}\frac{H_{\cD_{0}}^{2}}{H_{\cD}(z)^{2}}\right|\delta\Omega_{\CR}^{\cD_{0}}
  \nonumber \\
 &  & 
 +\left|\frac{1-(1+z)^{3}}{2}\frac{H_{\cD_{0}}^{2}}{H_{\cD}(z)^{2}}\right|\delta\Omega_{m}^{\cD_{0}}
\label{eq:dummy-4}\\
\frac{\delta r_{\perp}}{r_{\perp}} & = & 
\frac{\delta D_{M}}{D_{M_{0}}}=\frac{\delta H_{\cD_{0}}}{H_{\cD_{0}}}+
\left|\frac{I(z)^{2}}{6}+\frac{I^{\prime}(z)}{I(z)}\right|\delta\Omega_{\CR}^{\cD_{0}}
\nonumber \\
 &  & 
 +\left|\frac{I^{\prime}(z)}{I(z)}\right|\delta\Omega_{m}^{\cD_{0}}
\label{eq:dummy-5}\\
\frac{\delta D_{V}}{D_{V}} & = & \frac{1}{3}\frac{\delta r_{\parallel}}{r_{\parallel}}+\frac{2}{3}\frac{\delta r_{\perp}}{r_{\perp}}
\label{eq:dummy-6}
\end{eqnarray}
where $I^{\prime}(z)$ denotes a partial derivation
with respect to the respective parameter, i.e.~$\Omega_{\CR}^{\cD_{0}}$
or $\Omega_{m}^{\cD_{0}}$. Note that $I^{\prime}(z)$ and $I(z)$ are 
evaluated on the background
($\Omega_{\CR}^{\cD_{0}}=0$ and $\Omega_m^{\cD_{0}}=\Omega_m$).

\begin{figure}
\includegraphics[width=\linewidth]{%
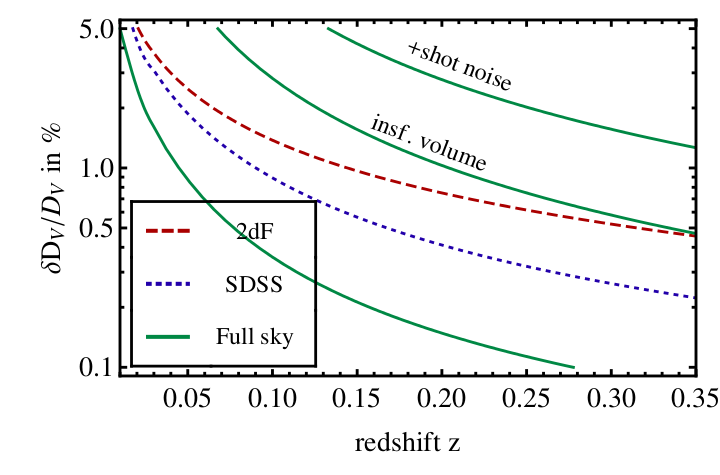}
\caption{
Errors on the distance $D_{V}$ for various survey geometries as a function of maximum redshift. For comparison the error induced by the finite number of BAO modes in the corresponding full sphere volume, calculated with the fitting formula of \cite{baoDistErrors}, is shown (insufficient volume). This error is about a factor of $10$ bigger than the error from the local volume 
distortion caused by inhomogeneities that we calculated. Adding a shot noise term, 
corresponding to a galaxy density of $n=3\times10^{-4} h^3 \mathrm{Mpc^{-3}}$ typical for 
SDSS and BOSS, we find that the cosmic variance of $D_V$ is a subdominant contribution to the error 
budget.
\label{fig:dDVcomp}}
\end{figure}

We evaluated the magnitude of the fluctuations in $D_V$, based on the cosmological 
parameters of the concordance model,  
as presented in Fig.~\ref{fig:dDVcomp}. Fluctuations as low as one per cent are 
reached for much smaller domains than for the cosmic variance of the 
$\Omega$-parameters. Thus, at first sight it might seem that the BAO measurement 
of $D_V$ could essentially overcome the cosmic variance limit. Closer inspection of this 
result reveals that this is not the case. Indeed, the much smaller variation of the distance $D_V$ 
means that a precise knowledge of the distance measure $D_{V}$ does
not lead to an equivalently good estimate of the cosmic parameters.

Clearly, the systematic uncertainty that we calculated is only a minor effect compared with 
the errors intrinsic to the actual measurement of the acoustic scale, as a
comparison of the three solid (green) lines in Fig.~\ref{fig:dDVcomp} shows. The lowest one is the
fluctuation of the scale $D_{V}$ for full spheres of the corresponding size at different places in
the universe. It is therefore the possible local deformation caused by statistical over- or underdensities.
The possible precision of a measurement of $D_{V}$ by BAOs, however, also depends on 
the number of observable modes. This induces an error if the volume is too small, and in 
particular when it is smaller than the BAO scale a reasonable measurement is no longer 
possible. Accordingly, even for a perfect sampling of the observed volume,  the error will not be smaller 
than the solid (green) lines in the middle. If one adds shot noise caused by imperfect sampling by 
a galaxy density of $n=3\times10^{-4} h^3 \mathrm{Mpc^{-3}}$, typical for SDSS and BOSS, 
the error increases even more. This means that for the realistic situation where we do not have 
a sufficiently small perfect ruler to allow for large statistics already for the small volumes considered here, the deformation uncertainty that we 
calculated remains completely subdominant.

\section{Fluctuations of the Hubble scale
\label{sec:Expansion-history}}

Local fluctuations of the Hubble expansion rate have already been considered 
in the literature 
\citep{hubble:turner,hubble:shiturner,hubble:wangturner,clarkson:hubble}. 
Here we wish to add two new aspects.

\begin{figure}
\includegraphics[width=\linewidth]{%
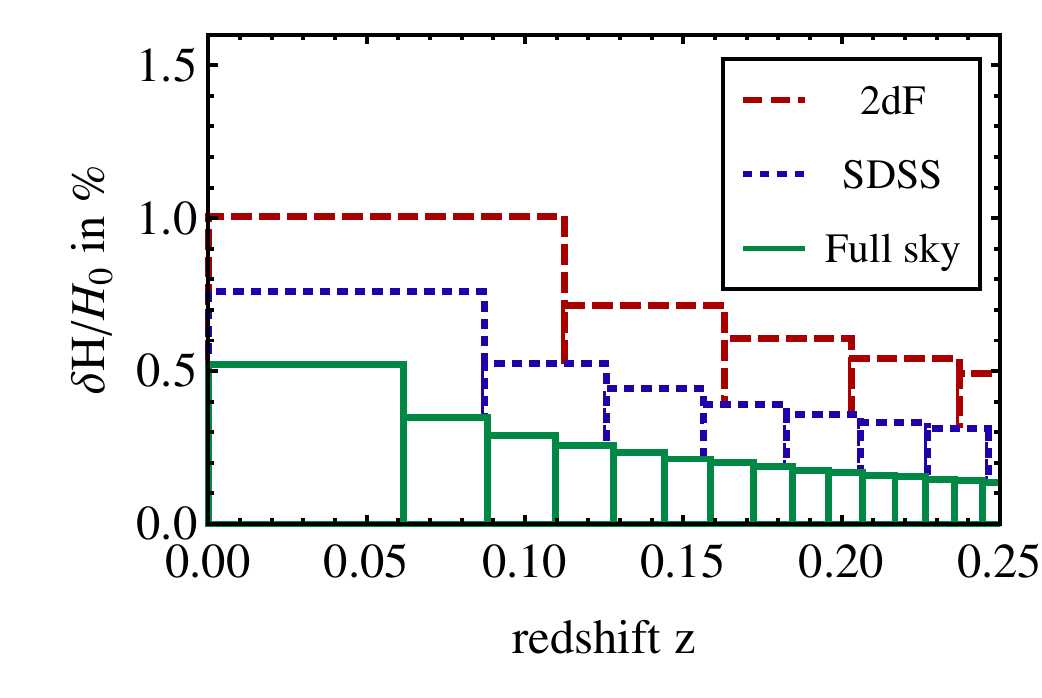}
\caption{
Optimum thickness
of shells to minimize the variance of $H\left(z\right)$ (see text
for the two competing effects). The respective error corresponds to
the height of the bars, the shells necessary for this purpose to their
width.
\label{fig:VarVerg}}
\end{figure}

The first one is on the measurement of $H(z)$ itself.
Experiments that try to measure $H$ as a function of
$z$, like the WiggleZ survey \citep{hubbleofz}, do this by measuring a ``local''
average $H(z_{m})$ in a region around the redshift $z_{m}$. These regions
should not be too small to keep the effects of local fluctuations
small. On the other hand they cannot be enlarged in an arbitrary
way because then the redshift $z_{m}$ becomes less and less characteristic
for the averaging domain. In other words, for an increasingly thicker
shell $\Delta z$, the evolution of $H(z)$ begins to play
a role. Therefore, one may find the optimal thickness of the averaging
shells over which the variation in the expansion rate
\begin{eqnarray}
{\rm Var}\left[H(z)\right] & = & \frac{1}{V_{\cD}}\intop H[z(r)]^{2}W_{\cD}\left(r\right){\rm d}^{3}r\nonumber \\
 &  & -\left(\frac{1}{V_{\cD}}\intop H[(z(r)]W_{\cD}\left(r\right){\rm d}^{3}r\right)^{2}
\label{eq:dummy}
\end{eqnarray}
equals the variance imposed by the inhomogeneous matter distribution.
The corresponding shells are shown in Fig.~\ref{fig:VarVerg}.
It should be noted that the error for the first
bin is certainly underestimated in our treatment, which rests on linear
perturbation theory. Taking into account higher orders, which become
dominant at small scales, will certainly increase it. Of course, in
these measurements the survey geometries will not necessarily be close
to the SDSS or the 2dF geometry, but they are shown to illustrate
survey geometries that do not cover the full sky.

Secondly, we wish to note that the relation between fluctuations in the Hubble expansion rate 
and fluctuations in the matter density offers the interesting possibility to determine the
evolution of the growth function for matter perturbations from the
variances of the Hubble rate measured at different redshifts. A direct measurement of the 
growth function by a determination of $\sigma_{8}$ at different epochs is difficult,
because one never examines the underlying dark matter distribution. Therefore one
has to assume that the observed objects represent 
the same clustering pattern as the underlying dark matter (this is the problem of bias).
It is well known that there is bias and its modeling typically has to rely on assumptions.

An interesting bypass is to look at the variation of local expansion rates at different redshifts. 
The assumption that the luminous objects follow the local flow is
more likely and the assumption that this local flow is generated by
the inhomogeneities of the underlying dark matter distribution is also reasonable. 
A similar idea leads to the attempt to use redshift-space distortions to do so 
\citep{percival:bao}.
The fact that one considers fluctuations means that we would not have
to know the actual value of $H\left(z\right)$, but only the local
variation at different redshifts.

This variation, defined as 
\begin{equation}
\delta_{H}=\frac{H_{\cD}-\overline{H}_{\cD}(a_{\cD})}{\overline{H}_{\cD}(a_{\cD})}\;,
\label{eq:vardH}
\end{equation}
has the fluctuations of Eq.~(\ref{eq:sigH})
\begin{equation}
\sigma\left(\delta_{H}\right)=\frac{1}{3}\overline{H_{\cD}}\left(a_{\cD}\right)f_\cD\left(a_{\cD}\right)\sigma_{\cD_{0}}\;.
\end{equation}
If we were to measure this quantity at different redshifts, we could,
without knowledge of the absolute normalization of $H_{\cD}\left(z\right)$, determine $f_\cD\left(a_{\cD}\right)$
only from the variance
and therefore the constant $c=\Omega_{\Lambda}/\Omega_{m}$.

Note that in the standard case, where the background redshift is 
identified with the observed one,
$f_\cD\left(a_{\cD}\right)$ is simply replaced by the growth rate $f\left(a\right)=\frac{\rmd \ln D\left(a\right)}{\rmd \ln a}$, and measuring the Hubble fluctuations would yield a direct measurement
of $f$. In the real world, where the redshift captures the structure on the way from the source to us,
it is not directly the background redshift. One would rather measure the modified 
"growth rate" $f_\cD \left(a_{\cD}\right)$.
The difference between these two quantities is  
small in our range of validity for $f_\cD\left(a_{\cD}\right)$, however (corrections of linear order in the perturbations).

\section{Conclusion
\label{sec:Conclusion}}

For the first time, we brought together the well--established perturbative 
approach to incorporate inhomogeneities in 
Friedman-Lema\^itre models (the theory of cosmological perturbations)
and the ideas of cosmological backreaction and cosmic averaging in the
Buchert formalism. Focusing on observations of the large--scale structure  
of the Universe at late times, we showed that the cosmic variance of
cosmological parameters is in fact the leading order contribution 
of cosmological averaging. 

We studied volume averages, their expected means, and variances 
of the cosmological parameters $H_0, \Omega_\CR, \Omega_m,
\Omega_\Lambda$ ($\Omega_\CQ$ is of higher order in
perturbation theory). Our central result is 
summarized in (\ref{eq:sigH})--(\ref{eq:sigOQ}).

Our extension of the backreaction study of \cite{li:scale} 
to the $\Lambda$CDM case enabled us to study fluctuations for a wider 
class of cosmological models. We were able to confirm for the fluctuations
in the Hubble rate that our results in comoving synchronous gauge agree
with those found in Poisson gauge
\citep{clarkson:hubble}.
 
The use of general window functions allowed us to consider realistic 
survey geometries in detail and to calculate the
fluctuations in the matter density, empirically found in the SDSS data by 
\cite{driver:cosvar}, directly from the underlying DM power spectrum. 
Converting this information into curvature fluctuations, we found that 
regions of $540 h^{-1} \mathrm{Mpc}$ diameter may still have a curvature 
parameter of $\sim 0.01$, even if the background curvature vanishes
exactly. We found that cosmic variance is a limiting factor even for surveys of the size 
of the SDSS survey. A volume--limited sample up to a redshift of $0.5$ 
was able to constrain the local curvature to $0.1$ per cent. 

Finally, we investigated the distortions of the local distance
to a given redshift and found that it is less affected by the fluctuations
of the local cosmic parameters than one might expect. The distance
measure $D_{V}$, used in BAO studies, is accurate to $0.2$ percent 
for samples ranging up to $z\approx0.35$. This means that BAO studies
are not limited by cosmic variance, but by problems such as sampling
variance and insufficient volume, as discussed in section~\ref{sec:Variances-on-BAO}.

In a next step one should incorporate the second--order effects into the
expected means of the cosmological parameters. There are no second--order
corrections to the variances, as argued in section~\ref{sec:Inhomogeneity-and-expansion}. Therefore a complete
second--order treatment seems feasible. 

The limitation of our approach comes from the fact that Buchert's
formalism relies on spatial averaging. Averaging on the light cone would
be more appropriate \citep{lightcone}, 
thus the study in this work has been restricted to 
redshifts $\ll 1$, where we expect light cone effects to play a
subdominant role.

\begin{acknowledgements}
We thank Thomas Buchert, Chris Clarkson, Julien Larena, Nan Li, Giovanni Marozzi, 
Will Percival, and Marina Seikel for discussions and valuable comments. 
We acknowledge financial support by Deutsche Forschungsgemeinschaft (DFG) 
under grant IRTG 881.
\end{acknowledgements}

\end{document}